\documentclass[showpacs,aps,prd,nofootinbib,floatfix,amsmath,amssymb]{revtex4}
\usepackage{graphicx}
\begin{document}

\makeatletter
%Feynman slash
\newbox\slashbox \setbox\slashbox=\hbox{$/$}
\newbox\Slashbox \setbox\Slashbox=\hbox{\large$/$}
\def\pFMslash#1{\setbox\@tempboxa=\hbox{$#1$}
  \@tempdima=0.5\wd\slashbox \advance\@tempdima 0.5\wd\@tempboxa
  \copy\slashbox \kern-\@tempdima \box\@tempboxa}
\def\pFMSlash#1{\setbox\@tempboxa=\hbox{$#1$}
  \@tempdima=0.5\wd\Slashbox \advance\@tempdima 0.5\wd\@tempboxa
  \copy\Slashbox \kern-\@tempdima \box\@tempboxa}
\def\FMslash{\protect\pFMslash}
\def\FMSlash{\protect\pFMSlash}
\def\miss#1{\ifmmode{/\mkern-11mu #1}\else{${/\mkern-11mu #1}$}\fi}
%%%% Uso:  \pFMSlash{p}
\makeatother

%\tightenlines
\title{Bounding the top and bottom electric dipole moments from neutron experimental data}
\author{A. Cordero--Cid}
\affiliation{Facultad de Ciencias de la Electr\' onica,
Benem\'erita Universidad Aut\'onoma de Puebla, Blvd. 18 Sur y Av.
San Claudio, 72590, Puebla, Pue., M\'exico.}
\author{J. M. Hern\' andez, G. Tavares--Velasco, and J. J. Toscano}
\affiliation{Facultad de Ciencias F\'{\i}sico Matem\'aticas,
Benem\'erita Universidad Aut\'onoma de Puebla, Apartado Postal
1152, Puebla, Pue., M\'exico.}

\begin{abstract}
Heavy quarks, namely, the top and bottom quarks, may show great
sensitiveness to new physics effects. In particular, they might have
unusually large electric dipole moments. This possibility is
analyzed via the corresponding one-loop correction to the neutron
electric dipole moment, $d_n$. The current experimental limit on
$d_n$ is used then to derive the uppers bounds $|d_t|<3.06\times 10^{-15}$ e-cm,
$|d_b|<1.22\times 10^{-13}$ e-cm.
\end{abstract}


\maketitle

The electric dipole moment (EDM) of elementary particles is a clear
signal of CP violation. Even more, such an electromagnetic property
would constitute itself a clear evidence of
beyond-the-standard-model (SM) CP-violating effects due to the large
suppression of the respective SM predictions. It is a well known
fact that the only source of CP violation in the SM, namely, the
Cabbibo--Kobayashi--Maskawa (CKM) phase, has an negligible effect on
flavor--diagonal processes such as the EDM of elementary particles
\cite{PR}. For instance, the EDM of quarks arises up to the
three--loop level \cite{SMEDMQ}. While the EDM of light fermions has
been long studied both theoretically and experimentally (the EDM of
light quarks via the neutron and proton) those of the heaviest
quarks still require more attention. In fact, the top quark may be
more sensitive to new sources of CP violation since it is the only
known fermion with a mass of the size of the electroweak symmetry
breaking scale. Indeed, most of the theories that predict new physics
effects beyond the Fermi scale, predict also EDMs several orders of
magnitude larger than those predicted by the SM.

\begin{figure}
\centering
\includegraphics[width=3.0in]{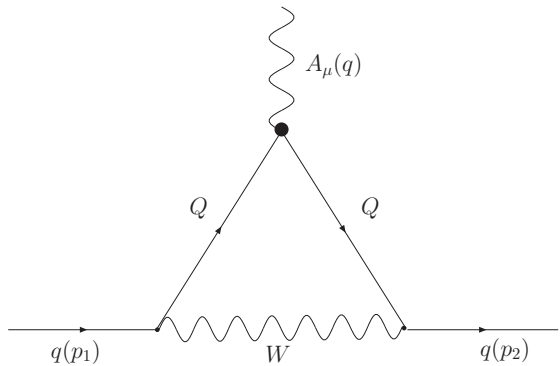}
\caption{\label{FIG1}Diagram contributing to the on-shell
$\bar{q}q\gamma$  vertex. $Q$ stands for a heavy quark and $q$ for a
light one.}
\end{figure}

In this note, we will analyze the effects induced by the EDM
of the $t$ and $b$ quarks on the  one--loop-induced EDM of the $d$
and $u$ quarks. We will use then the experimental limit on the
neutron EDM to constrain the one associated with the $t$ and $b$
quarks. The EDM of heavy quarks, from now on denoted by $Q$, can be
parametrized by the following Lagrangian
\begin{equation}
{\cal L}_{QQ\gamma}=-\frac{i}{2}d_Q\bar{Q}\gamma_5\sigma_{\mu
\nu}QF^{\mu \nu},
\end{equation}
where $d_Q$ stands for the $Q$ quark EDM. In the unitary gauge, the
contribution to the on--shell $\bar{q}q\gamma$ coupling is given
through the diagram shown in Fig. \ref{FIG1}. The respective
one--loop vertex can be written as
\begin{equation}
\label{am}
\Gamma_\mu=-\frac{g^2|V_{Qq}|^2d_Qm_Q}{2}\int\frac{d^4k}{(2\pi)^4}\frac{P_R\gamma_\alpha
[(\pFMSlash{k}-\pFMSlash{p_2})\sigma_{\mu \nu}q^\nu-\sigma_{\mu
\nu}q^\nu (\pFMSlash{k}-\pFMSlash{p_1})]\gamma_\beta P^{\alpha
\beta}}{\left(k^2-m^2_W\right)\left((k-p_1)^2-m^2_Q\right)\left((k-p_2)^2-m^2_Q\right)},
\end{equation}
where $Q=t$ or $b$, $q=u$ or $d$, and $V_{Qq}$ is the associated CKM
element. In addition,
\begin{eqnarray}
P^{\alpha \beta}&=&g^{\alpha \beta}-\frac{k^\alpha k^\beta}{m^2_W}.
\end{eqnarray}
Below we will ignore the longitudinal component of the $W$ gauge
boson owing to the fact that it contributes marginally to the above
amplitude. This is a good approximation indeed as the dropped terms
are proportional to increasing powers of $(m_q/m_W)^2$, which in
fact is a negligible quantity. Once this approximation is taken into
account, which greatly simplifies the calculation, we are left with
the following term
\begin{equation}
\Gamma_\mu=-\frac{g^2|V_{Qq}|^2d_Qm_Q m_q}{2}\int\frac{d^4k}{(2\pi)^4}\frac{P_R[\sigma_{\mu
\nu}q^\nu-2i(k-p_1)_\mu]}{\left(k^2-m^2_W\right)\left((k-p_1)^2-m^2_Q\right)\left((k-p_2)^2-m^2_Q\right)}.
\end{equation}
Notice that there are contributions to both the magnetic and
electric dipole moments of the $q$ quark, but we are only interested
in its CP--odd property.  The contribution proportional to
$p_{1\mu}$ can be expressed in terms of $\sigma_{\mu \nu}q^\nu$ via
Gordon's identity, whereas the part proportional to $\gamma_\mu$ can
be ignored as it is proportional to $m_q$. Notice that gauge
invariance is preserved at this order. After evaluating the integral
over $k$, we obtain
\begin{equation}
d_q=\Big(\frac{\alpha}{4\pi}\Big)\Bigg(\frac{|V_{Qq}|^2}{s^2_W}\Bigg)d_Q\sqrt{x_qx_Q}f(x_q,x_Q),
\end{equation}
where $s_W=\sin\theta_W$, $x_a=m^2_a/m^2_W$ and $f(x_q,x_Q)$ stands
for the following integral
\begin{equation}
f(x_q,x_Q)=\int^1_0
dx\int^{1-x}_0dy\frac{2-x-y}{1-(1-x_Q)(x+y)-x_q(1-x-y)(x+y)}.
\end{equation}
Although this integral has analytical solution in the most general
case, a compact solution is found in the $x_q=0$ approximation:
\begin{equation}
f(x_Q)=\frac{1+3x_Q(x_Q-4/3)-2(2x_Q-1)\log(x_Q)}{2(x_Q-1)^3}.
\end{equation}

We turn now to express the neutron EDM, $d_n$, in terms of the ones
associated with its constituents, $u$ and $d$. We will use the
non-relativistic $SU(6)$ formula, which  holds at the electro-weak
scale. One then invokes renormalization group to evolve the electric
dipole operator down to the hadronic scale. The neutron EDM at the
hadronic scale is thus given by

\begin{equation}
d_n=\eta^E\left(\frac{4}{3}d_d-\frac{1}{3}d_u\right)
\end{equation}
where $\eta^E\sim 0.61$ is the QCD correction factor. We can thus write
\begin{equation}
|d_n|=\Bigg|\eta^E\left(\frac{\alpha}{4\pi s_W^2}\right)\Bigg[\frac{4}{3}\sqrt{x_t\,x_d}|V_{td}|^2f(x_t)d_t-
\frac{1}{3}\sqrt{x_bx_u}|V_{bu}|^2f(x_b)d_b\Bigg]\Bigg|.
\end{equation}
For consistency with our analysis, we will use the current quark
masses for the $u$ and $d$ quarks. Taking into account that $f(x_t)=0.22$ and $f(x_b)=5.07$, numerical evaluation gives:
\begin{equation}
|d_n|=|9.48 \times 10^{-12}d_t-2.37\times 10^{-13}d_b|
\end{equation}
We are now able to constrain the $t$ and $b$ quarks EDMs. The experimental limit on the neutron EDM is given by
\cite{Baker:2006ts}:
\begin{equation}
d^{\rm Exp}_n<2.9\times 10^{-26}\ {\rm e\cdot cm}.
\end{equation}
As far as the CKM matrix elements is concerned, the latest reported values are \cite{PDG}:
\begin{eqnarray}
|V_{td}|&=&(7.40\pm 0.80)\times 10^{-3}, \\
|V_{bu}|&=&(4.31\pm 0.30)\times 10^{-3}.
\end{eqnarray}

To obtain a constraint on the EDM of the $t$ and $b$ quarks, we will assume that
either $d_t$ or $d_b$ is nonzero and the other one vanishes.  We then obtain
\begin{eqnarray}
|d_t|&<&3.06\times 10^{-15}\ {\rm e\cdot cm}, \\
|d_b|&<&1.22\times 10^{-13}\ {\rm e\cdot cm}.
\end{eqnarray}

It is worth comparing these results with theoretical expectations. The electromagnetic dipolar
structure of the top quark has been studied in diverse contexts. We will present a brief survey of the theoretical expectations for the heavy quark EDMs and the corresponding experimental bounds on them. As already mentioned, quark EDM arises in the SM up to the three-loop level \cite{SMEDMQ}. The CP-violating effects arise from the CKM phase. From the estimate for the three-loop contribution for the $d$ quark EDM, one can roughly estimate that $d_t$ is of the order of $10^{-31}$ to $10^{-32}$ e-cm. In contrast, in some SM extensions, such as supersymmetric theories and multi-Higgs doublet models (MHDMs) the situation is quite different as fermion EDM can arise at the one-loop level \cite{MHM,Soni:1992tn,Bernreuther,Atwood:1995,susy,Iltan:2004xr}, with the corresponding estimate for $d_t$ being about ten orders of magnitude larger than the SM prediction. It has been noted that this result opens the window for experimental detection \cite{Atwood:2000tu}. Such large values for the $t$ quark EDM are due to the fact that in models with an extended Higgs sector the fermion EDM scales as $m_f^3$.  Heavy quark EDM has been calculated extensively in the framework of MHDMs, where CP violation can arise due to the scalar exchange between quarks. It has been found that \cite{Soni:1992tn}, assuming that the dominant contribution arises from the lightest neutral Higgs boson $h$, $d_t$ can be of the order of $10^{-19}$ e-cm for $m_h=100$ GeV. On the other hand, CP violation can also arise in the neutral Higgs sector of a two-Higgs doublet model (THDM) if there is a phase in the Higgs-fermion-fermion interactions. This scenario has been explored in Ref \cite{Bernreuther}. Although this analysis refers to the type II
THDM, it is also valid for type I and III THDMs after some replacements of the coupling constants. For particular values of the parameters of THDM with CP violation in the neutral scalar sector, the $t$ quark EDM can reach values of the order of $10^{-18}$-$10^{-19}$ for $m_h= 100-300$ GeV \cite{Atwood:2000tu}. Even more, in models with two or three Higgs doublets, CP violation can also arise in the charged Higgs sector. Assuming $m_{H^\pm}=200$ GeV, authors of Ref. \cite{Atwood:1995} showed that $d_t$ is of the order of $10^{-22}$. As for supersymmetric theories, $t$ quark EDM can arise at the one-loop level in the minimal supersymmetric standard model (MSSM) even without generation mixing. The CP-violating phase is provided by the chargino and neutralino mixing matrices as well as the squarks $q_L - q_R$ mixing matrices. In this model the $d_t$ can receive gluino, chargino and neutralino contributions \cite{susy}. For convenient values of the parameters involved in the calculation, the $t$ quark EDM is typically of the order of $10^{-19}$-$10^{-20}$ e-cm, which is much smaller than the values expected in the Higgs sector. The EDM of heavy quarks is also sensitive to non-universal extra dimensions \cite{Iltan:2004xr}. It has been estimated that the $t$ quark EDM can reach values as high as $10^{-20}$ e-cm for a value of the compactification scale of $300$ GeV. A more
suppressed value for $d_t$, of the order of $10^{-22}$ e-cm, was
obtained recently from the one--loop contribution of an anomalous
$tbW$ vertex including both left-- and right--handed complex
components \cite{TT}, and more recently the one--loop contribution of
an anomalous CP--odd $WW\gamma$ vertex \cite{NT} was used to
estimate the value of $10^{-21}$ e-cm for $d_t$.

On the experimental side, the data on the rare flavor changing
neutral current decay $b\to s\gamma$ has been used to analyze
potential top--mediated new physics effects \cite{HR,T}. In Ref.
\cite{HR} the contribution to the $b\to s\gamma$ decay from both the
magnetic and electric dipole moments of the top quark was studied.
Those authors report an  upper bound on $d_t$ of the order of
$10^{-16}$ e-cm, which is stronger than that
obtained here. Detailed studies have also been made to probe the structure of the $tt\gamma$ vertex at future $e^-e^+$, $pp$, $p\bar{p}$, $\gamma\gamma$ and $\mu^-\mu^+$ colliders. Although there is no doubt that it is of extreme importance the measurement of the static EDM of the $t$ quark, due to its short lifetime it will be much less difficult to measure the $t$ quark nonstatic EDM. One of the main tasks of the future linear $e^-e^+$ collider will be to determine the top quark properties, mainly via $t\bar{t}$ production. It has been found \cite{Atwood:1992} that it will be possible to determine values of the order of $d_t\sim 10^{-17}$ e-cm with $10^4$ $t\bar{t}$ events in a linear collider running at c.m. energies of $\sqrt{s} = 500$ GeV. The $t$ quark dipole moment could also be probed at a future photon collider via $\gamma\gamma\to t\bar{t}$ \cite{Choi:1995kp}. It has been found that for c.m. energies of 500 GeV, a photon collider would be sensitive to values of the order of $d_t\sim 10^{-17}$ e-cm. Therefore, the limits on the $t$ quark EDM that would be obtained at a photon collider are of the same order than those that would be obtained at an $e^-e^+$ linear collider.

As far as the $b$ quark EDM is concerned, values which differ by one or two orders of magnitude than those obtained for the $t$ quark EDM have been reported in the literature.  To our knowledge there is not yet any upper
bound reported in the literature, but there are estimates derived
from multi--Higgs models. In this class of theories, the
estimates for $d_b$ are reported to lie in the range of
$10^{-23}-10^{-22}$ e-cm, whereas the  estimate
$10^{-23}$\ e-cm was derived from the one--loop contribution of a
CP--odd $WW\gamma$ vertex \cite{NT}.

In conclusion, we have used the current experimental limit on the neutron EDM to
derive upper bounds on the $t$ and $b$ quarks EDMs. Our constraint on $d_t$ is weaker than that derived from the $b\to s\gamma$
decay \cite{HR}. Roughly speaking, the bounds on $d_t$ and $d_b$ are
at least four orders of magnitude above the predictions obtained in
several specific models incorporating new sources of CP violation.
It means that there is a potential window to explore CP--violating
effects in the third quark family.

\acknowledgments{We acknowledge financial support from CONACYT
(M\' exico).}


\begin{thebibliography}{99}

\bibitem{PR}For a recent review, see M. Pospelov and A. Ritz,
Ann. Phys. (N.Y.) \textbf{318}, 119 (2005).

\bibitem{SMEDMQ} M. E. Pospelov and I. B. Khriplovich, Sov. J.
Nucl. Phys. \textbf{53}, 638 (1991) [Yad. Fiz. \textbf{53}, 1030
(1991)]; E. P. Shabalin, Sov. J. Nucl. Phys. \textbf{28}, 75
(1978) [Yad. Fiz. \textbf{28}, 151 (1978)]; D. Chang, W. Y. Keung,
and J. Liu, Nucl. Phys. \textbf{B355}, 295 (1991);
A. Czarnecki, B. Krause, Phys. Rev. Lett. 78 (1997) 4339;
I. B. Khriplovich, Phys. Lett. B 173 (1986) 193.

\bibitem{Baker:2006ts}
  C.~A.~Baker {\it et al.},
  %``An improved experimental limit on the electric dipole moment of the
  %neutron,''
  Phys.\ Rev.\ Lett.\  {\bf 97}, 131801 (2006)
  [arXiv:hep-ex/0602020].
  %%CITATION = PRLTA,97,131801;%%

\bibitem{PDG} W. -M. Yao,  \textit{et al.} (Particle Data Group),
J. Phys. \textbf{G33}, 1 (2006).


\bibitem{MHM} S. Weinberg, Phys. Rev. Lett. \textbf{58}, 657
(1976); G. C. Branco and M. N. Robelo, Phys. Lett. \textbf{B160},
117 (1985); J. Liu and L. Wolfenstein, Nucl. Phys. \textbf{B289},
1 (1987); C. H. Albright, J. Smith, and S. H. H. Tye, Phys. Rev.
\textbf{D21}, 711 (1980);  N. G. Deshpande and E. Ma, Phys. Rev.
\textbf{D16}, 1583 (1977); Y. Liao and X. Li, Phys. Rev.
\textbf{D60}, 073004 (1999); D. G. Dumm and G. A. Sprinberg, Eur.
Phys. J. \textbf{C11}, 293 (1999); D. A. Demir and M. B. Voloshin,
Phys. Rev. \textbf{D63}, 115011 (2001); E. O. Iltan, J. Phys.
\textbf{G27}, 1723 (2001); Phys. Rev. \textbf{D65}, 073013 (2002).

\bibitem{Soni:1992tn}
  A.~Soni and R.~M.~Xu,
  %``Electric dipole moment of the top quark on CP nonconserving Higgs boson
  %exchange,''
  Phys.\ Rev.\ Lett.\  {\bf 69}, 33 (1992).

\bibitem{Bernreuther}W. Bernreuther, T. Schroder and T.N. Pham, Phys. Lett. B279, (1992) 389; 
C.D. Froggatt, R.G. Moorhouse and I.G. Knowles, Nucl. Phys. B386, (1992) 63.

\bibitem{Atwood:1995}D. Atwood, S. Bar-Shalom and A. Soni, Phys. Rev. D51, (1995) 1034.

\bibitem{susy}E. Christova and M. Fabbrichesi, Phys. Lett. B315, (1993) 338; W. Bernreuther and P. Overmann, Z. Phys. C61, (1994) 599; B. Grzadkowski, 
Phys. Lett. B305, (1993) 384; A. Bartl, E. Christova and W. Majerotto, Nucl. Phys. B460, (1996) 235; Erratum-ibid.
Nucl. Phys. B465, (1996) 365; A. Bartl, E. Christova, T. Gajdosik and W. Majerotto, Nucl. Phys. B507, (1997) 35;
Erratum-ibid. B531, (1998) 653.

%\cite{Iltan:2004xr}
\bibitem{Iltan:2004xr}E.~O.~Iltan, JHEP {\bf 0404}, 018 (2004).

%\cite{Atwood:2000tu}
\bibitem{Atwood:2000tu}
  D.~Atwood, S.~Bar-Shalom, G.~Eilam and A.~Soni,
  %``CP violation in top physics,''
  Phys.\ Rept.\  {\bf 347}, 1 (2001)

\bibitem{TT} J. Hern\' andez--S\' anchez {\it et al.}, Phys. Rev.
\textbf{D75}, 073017 (2007).

\bibitem{NT} H. Novales--S\' anchez and J. J. Toscano, work in
progress.

\bibitem{HR} J. L. Hewett and T. G. Rizzo, Phys. Rev.
\textbf{D49}, 319 (1994).

\bibitem{T} R. Mart\'\i nez, M. A. P\' erez, and J. J. Toscano,
Phys. Lett. \textbf{B340}, 91 (1994).

\bibitem{Atwood:1992}D. Atwood and A. Soni, Phys. Rev. D45 (1992) 2405; W. Bernreuther and P. Overmann, Z. Phys. C72, (1996) 461; 
W. Bernreuther, A. Brandenburg and P. Overmann, hep-ph/9602273, in {\it $e^+e^-$ Linear
Collisions: The Physics Potential}, P. 49, P. M. Zerwas ed. (1995) page 49.

%\cite{Choi:1995kp}
\bibitem{Choi:1995kp}
  S.~Y.~Choi and K.~Hagiwara,
  %``Probing the top quark electric dipole moment at a photon linear collider,''
  Phys.\ Lett.\  B {\bf 359}, 369 (1995); M.S. Baek, S.Y. Choi and C.S. Kim, Phys. Rev. D56, (1997) 6835.



\end{thebibliography}
\end{document}